\documentclass[epj]{svjour}
\usepackage[]{color}
\usepackage[]{graphicx}
\usepackage[]{enumerate}
\usepackage[]{lineno}%\linenumbers
\usepackage[latin1]{inputenc}  %\"umläute
\usepackage[T1]{fontenc}
\usepackage{url}

\graphicspath{{../Figures-EPS/}{.}}
\begin{document}
\title{Efficiency determination of resistive plate chambers for fast quasi--monoenergetic neutrons}

\author{M.~R\"{o}der\inst{1,2} \and Z.~Elekes\inst{3,1} \and T.~Aumann\inst{4,5} \and D.~Bemmerer\inst{1}\thanks{d.bemmerer@hzdr.de} \and
K.~Boretzky\inst{4} \and C.~Caesar\inst{4,5} \and 
T.E.~Cowan\inst{1,2} \and 
J.~Hehner\inst{4} \and % Konstanze 16.04.
M.~Heil\inst{4} \and % Konstanze 16.04.
M.~Kempe\inst{1,2} \and V.~Maroussov\inst{6,4} \and O.~Nusair\inst{4,7}\thanks{Present address: Argonne National Laboratory, Argonne, Illinois, USA} \and A.~V.~Prokofiev\inst{8} \and 
R.~Reifarth\inst{9} \and % Konstanze 16.04.
M.~Sobiella\inst{1} \and 
D.~Stach\inst{1} \and 
A.~Wagner\inst{1} \and D.~Yakorev\inst{1,2} 
\and A.~Zilges\inst{6} % Konstanze 16.04.
\and K.~Zuber\inst{2} for the R$^3$B collaboration
}                     % Do not remove
\institute{%
Helmholtz-Zentrum Dresden-Rossendorf, Dresden, Germany \and
Technische Universit\"{a}t Dresden, Dresden, Germany \and
MTA ATOMKI, Debrecen, Hungary \and
GSI Helmholtzzentrum f\"{u}r Schwerionenforschung, Darmstadt, Germany \and
Technische Universit\"{a}t Darmstadt, Darmstadt, Germany \and
Universität zu K\"{o}ln, K\"{o}ln, Germany \and
Al-Balqa Applied University, Salt, Jordan \and
The Svedberg Laboratory, Uppsala University, Uppsala, Sweden \and
Johann Wolfgang Goethe -- Universität, Frankfurt am Main, Germany 
}
\date{\today}
\abstract{
Composite detectors made of stainless steel converters and multigap resistive plate chambers have been irradiated with quasi-monoenergetic neutrons with a peak energy of 175\,MeV. The neutron detection efficiency has been determined using two different methods. The data are in agreement with the output of Monte Carlo simulations. The simulations are then extended to study the response of a hypothetical array made of these detectors to energetic neutrons from a radioactive ion beam experiment.
}
 % end of PACS codes
\maketitle
\section{Motivation}
\label{intro}
Nuclear reactions involving nuclei close to or beyond the neutron drip line must be studied in order to understand the synthesis of the heavy chemical elements \cite{Wiescher12-ARAA}. One of the experimental methods to investigate such exotic nuclear systems
is the invariant mass method, which requires a kinematically complete measurement. By detecting
and identifying the products of the nuclear reaction in question and determining their momentum, the invariant mass of the system can be reconstructed. Due to
the abundance of  neutrons in the typical nuclei under investigation, this technique usually involves
neutron detection. 

The R$^{3}$B (Reactions with Relativistic Radioactive \linebreak Beams) collaboration aims to study the properties
of such exotic nuclei \cite{Aumann07-PPNP}. At the present R$^{3}$B setup in cave C of GSI, the Large Area Neutron Detector (LAND)~\cite{Blaich92-NIMA} is used. This device covers 2$\times$2\,m$^2$ area and reaches a single-neutron efficiency of 90\% for 0.5\,GeV neutrons, with a typical time-of-flight resolution of $\sigma$ = 250\,ps \cite{Boretzky03-PRC}. 
Similar but smaller detectors for high-energy neutrons exist at radioactive ion beam facilities in the United States \cite{Baumann05-NIMA} and Japan \cite{Kobayashi13-NIMB} and at the COSY-TOF spectrometer in Jülich, Germany \cite{Karsch01-NIMA}.

The ongoing construction of a new infrastructure for producing
radioactive ion beams named FAIR (Facility for Antiproton and Ion Research) in Darmstadt 
requires a more powerful neutron detector for the future R$^3$B setup, with the working title NeuLAND \cite{NeuLAND-TDR11}. 

The properties aimed for with NeuLAND are unprecedented for a fast neutron (0.2-1.0~GeV) array: at least 90\% detection efficiency for single neutrons at 0.2~GeV energy, 
a very large angular coverage of 80~mrad at a distance of 12~m to the reaction target,
an excellent time resolution of $\sigma$ = 150~ps and
the needed granularity to achieve good invariant mass resolution, e.g.,
$\sigma$=20 keV at 100 keV excitation energy above the threshold for medium mass systems.
Also, the setup must be able to identify multi-neutron
events with up to four neutrons per event, and correctly reconstruct their momentum.

There are two basic approaches in order to build such a device, and both have been studied in depth during the development phase of NeuLAND \cite{Aumann10-AnnRep}:

\begin{itemize}
\item[(A)]
A detector may be built out of a large, suitably subdivided volume of plastic scintillator material, exploiting the interaction of energetic neutrons with the protons included in the plastic for the detection \cite{Baumann05-NIMA,Kobayashi13-NIMB,Karsch01-NIMA}. 
\item[(B)] Alternatively, a hybrid approach can be chosen, with a converter material with relatively large density and atomic charge number (e.g. iron) efficiently converting neutrons to high-energy protons. The energetic protons, in turn, are subsequently detected in a charged-particle detector with excellent time resolution. Converter and detector layers must be alternated, so that the resolution is not limited by insufficient granularity. This approach has been used for the previous LAND detector \cite{Blaich92-NIMA}. For NeuLAND it was studied whether a similar approach with smaller granularity may be adopted, replacing the scintillator material with multigap resistive plate chamber (MRPC) detectors with their excellent time resolution and less expensive readout.
\end{itemize}

In the end, approach (A) was selected for NeuLAND, based on the better multi-hit capability of a fully active device \cite{NeuLAND-TDR11} when compared with a converter-based detector. However, detectors for energetic neutrons based on approach (B) may have applications in other fields, in particular where their more limited multi-neutron capability  \cite{Elekes13-NIMA} is sufficient and where construction and operating costs pose a challenge. 

In the present work, the results from an in-beam test of a neutron detector based on approach (B) are reported. That detector, developed at HZDR, was tested in a campaign together with similar counters developed at GSI \cite{Caesar12-NIMA}. The measurement campaign and the present work are part of the development effort of NeuLAND, which was recently concluded in a Technical Design Report \cite{NeuLAND-TDR11}. 

%%%%%%%%%%%%%%%%%%%%%%%%%%%%%%%%%%%%%%%%%%%%%%%%%%%%%%%%%%%%%%%%%%%%%%%%%%%%%%%%%%%%%%%%
\section{Description of the detector under study}
\label{sec:DetectorDescription}
%%%%%%%%%%%%%%%%%%%%%%%%%%%%%%%%%%%%%%%%%%%%%%%%%%%%%%%%%%%%%%%%%%%%%%%%%%%%%%%%%%%%%%%%
\begin{figure}[]
\centering
\includegraphics[width=\columnwidth]{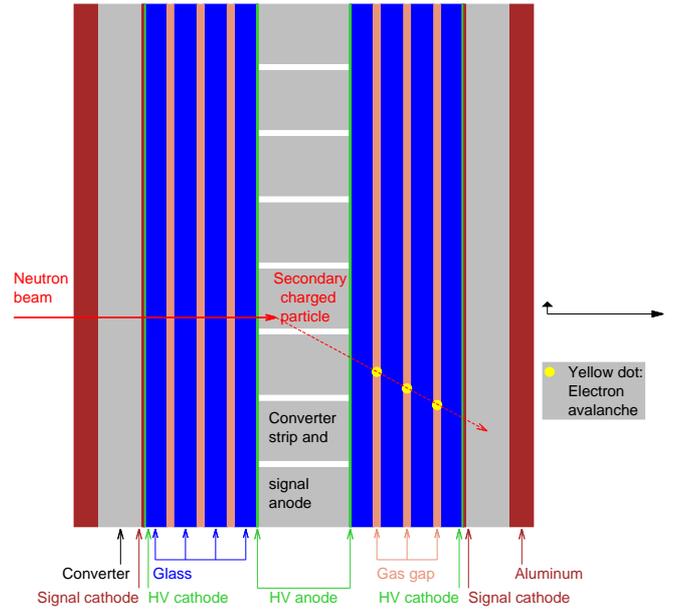}
\caption{Schematic view of the 2$\times$3\,gap MRPC based prototype under study here. The neutron beam impinges from the left. When a neutron is converted to a charged particle in the left converter plate or in one of the central converter strips, the charged particle is subsequently detected in the MRPC structure. The charge induced on the converter strips is amplified and used for the time and charge signal. The arrows on the right side of the plot are 5\,mm long each; note the different scale in x and y directions used for clarity.}
\label{fig:Prototype}
\end{figure}

The detector under study was built using an iron converter and multigap resistive plate chambers (MRPCs).
A resistive plate chamber is a gas-filled, parallel plate avalanche device, the electrodes of which are made from high-resistivity material. It was developed over 30 years ago \cite{Santonico81-NIM} and has since been used in a number of experiments \cite[for a review]{Peskov-RPC2012}. MRPCs are known for their excellent time resolution reaching $\sigma$ = 25\,ps~\cite{An08-NIMA}.

For the converter, stainless steel has been chosen. This material offers excellent structural stability in connection with a relatively high density. The impinging neutron beam passes, in sequence (fig.\,\ref{fig:Prototype}), 
\begin{itemize}
\item the 1\,mm thick aluminum housing of the prototype, 
\item then a 2\,mm thick converter plate, 
\item then the first half of the 2$\times$3\,gap MRPC structure, 
\item then 4\,mm thick, 25\,mm wide converter strips that are also used for the read out of the electrical signal (induced charge), 
\item then the second half of the 2$\times$3\,gap MRPC structure,
\item and finally a 2\,mm thick plate of converter material.
\end{itemize}
The MRPC structure consists of three gas gaps of 0.3\,mm thickness each. The working gas was a mixture of 85\% Freon R134a, 10\% SF$_6$, and 5\% iso-butane and was continuously flushed in order to avoid degradation of the detector performance. The gas gaps were separated by 1\,mm thick layers of float glass held in place by fishing lines. A typical electric field of 110\,kV/cm was applied between the outside of the structure (negative high voltage) and the center (ground potential), guaranteeing $>$95\% efficiency for minimum ionizing particles in the present 2$\times$3 gap structure. The high voltage electrodes used were made of semiconductive coated mylar. 

Neutrons hitting one of the converter plates produce charged particles. These charged particles generate an electron avalanche in the gas gaps, which, in turn, generates an induced charge on the central readout anode. The central converter, which was also used to read out the signals, was segmented with a strip width of 25\,mm and an interstrip spacing of 3\,mm. Two different prototypes called HZDR-1b and HZDR-3c, respectively, were used. Further details on the prototypes studied here can be found elsewhere \cite{Yakorev11-NIMA}. 

The detector design is intended to be iterated about 50 times in order to reach $>$90\% efficiency. In that scenario, the final 2\,mm plate and the 2\,mm entrance plate of the subsequent MRPC structure form together 4\,mm effective thickness. The prototype studied in the present work consists of just one 2$\times$3\,gap MRPC structure. 

During the development phase of the detector, prototypes of 40$\times$20~cm$^2$ (reduced size) and 200$\times$50~cm$^2$ (final size) were produced and tested.  The optimization process was carried out on the reduced size prototypes and involved numerous tests at the superconducting electron accelerator ELBE, which provided an electron beam of 30~MeV with time resolution of a few ps \cite{Yakorev11-NIMA}.

Both for the reduced size \cite{Yakorev11-NIMA} and for the final size \cite{Elekes13-NIMA} prototypes, it was proven by electron beam experiments at ELBE that the MRPC-based neutron detectors can deliver the requested $\sigma$ $\leq$ 100~ps time resolution and $\geq$90\% efficiency for minimum ionizing particles. Extensive Monte Carlo simulations were carried out \cite{Elekes13-NIMA} in order to extend these experimental findings to a simulated response for the full detector, which should include 50 subsequent layers of the final size prototype.

In order to further firm up the conclusions from the Monte Carlo simulations, several prototypes of reduced size were transported to the quasi-monochromatic neutron beam facility of The Svedberg Laboratory~(TSL)~\cite{Prokofiev07-RPD} at Uppsala University, Sweden. 

The neutron kinetic energy of 175\,MeV available at TSL is just below the lower edge of the  energetic range foreseen for NeuLAND, 0.2-1.0\,GeV. The performance of a converter-based detector should in principle improve with increasing kinetic neutron energy. This is so, because at higher neutron kinetic energy, secondary particles produced in the converter are more likely to have sufficient energy to reach the MRPC structure and are more forward focused, minimizing efficiency losses to the sides. The present experiment just below the lower energy range of NeuLAND is, therefore, suitable for a partial but stringent test of whether the working principle of the above described approach (B) for NeuLAND is sound.

%%%%%%%%%%%%%%%%%%%%%%%%%%%%%%%%%%%%%%%%%%%%%%%%%%%%%%%%%%%%%%%%%%%%%%%%%%%%%%%%%%%%%%%%
\section{Experiment at TSL Uppsala}
\label{sec:TSL}
%%%%%%%%%%%%%%%%%%%%%%%%%%%%%%%%%%%%%%%%%%%%%%%%%%%%%%%%%%%%%%%%%%%%%%%%%%%%%%%%%%%%%%%%
\begin{figure}[tb]
\centering
\includegraphics[width=\columnwidth]{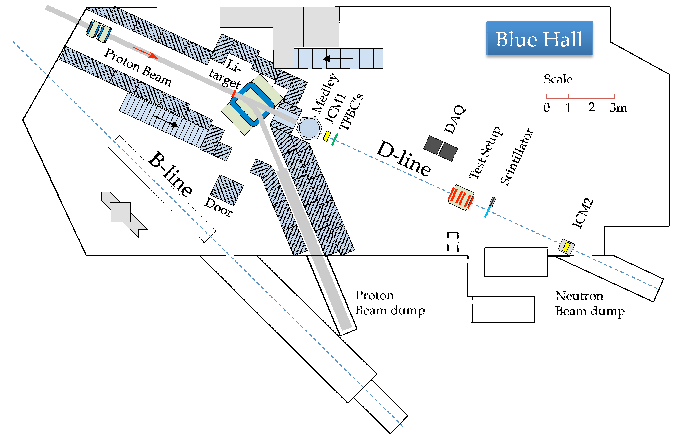}
\caption{Map of the TSL Uppsala quasi-monochromatic neutron irradiation facility. The $^7$Li(p,n)$^7$Be production target is just to the left of the bending magnet shown in blue. The neutron beam passes a long steel collimator. Then, it irradiates first the MEDLEY setup \cite{Bevilacqua11-NIMA}, subsequently the setup for the present experiment. }
\label{fig:TSL}
\end{figure}

The experiment took place at the quasi-monochromatic neutron irradiation facility in the "Blue Hall" of TSL Uppsala \cite{Prokofiev07-RPD}. A proton beam of 179~MeV hit a 23.5~mm thick target of metallic lithium, enriched to 99.99\% in $^7$Li (fig.\,\ref{fig:TSL}). The target had a rectangular shape of 20$\times$32~mm$^2$ and was directly water-cooled. The proton beam current on target was typically 300~nA.
Just downstream of the lithium target the protons were deflected by a magnet and
transported to a water-cooled beam dump made of graphite.
The neutron beam produced by the $^7$Li(p,n)$^7$Be reaction in the target
was then shaped successively by several collimators made of iron:
\begin{itemize}
\item cylindrical with a diameter of 20~cm and a length of 80~cm placed at 137~cm from the lithium target,
\item conical with a diameter changing from 30.09~cm to 47.48~cm and a length of 100~cm placed at 217~cm from the lithium target,
\item conical with a diameter changing from 57.11~cm to 61.46~cm and a length of 25~cm placed at 331~cm from the lithium target.
\end{itemize}
These collimators were surrounded by concrete blocks, which also served to shield the user area from the production target.

Just after the exit of the collimators, there was another user setup, MEDLEY~\cite{Bevilacqua11-NIMA}. MEDLEY exposed only a thin reaction target of typical thickness 1\,mm to the neutron beam. Thus, it attenuated the beam only negligibly, and no significant flux of disturbing particles is expected outside the MEDLEY chamber. 

Downstream of MEDLEY, the detectors used for the present experiment were arranged in the path of the neutron beam. The neutron flux was measured by two ionization chamber monitors (ICMs). The first ICM was placed just downstream of MEDLEY, at a distance of 5\,m from the lithium target, and the second one at the very end of the hall just before the neutron beam dump, at a distance of 16\,m from the lithium target. As an additional neutron flux monitor, altogether three thin film breakdown counters (TFBCs \cite{Prokofiev99-Techreport}) were used, all of them placed just downstream of the first ICM.

For the purposes of the present analysis, the neutron flux at the irradiation position was taken to be the weighted average of the fluxes determined from the first ICM (calibrated to 10\% precision), from a large TFBC (calibrated to 10\%), and from the current in the proton beam dump (neutron conversion coefficient calibrated to 30\%), all three fluxes converted to an effective flux at the MRPC position, properly taking the distances into account. The neutron fluxmeters were calibrated for the main, quasi-monochromatic peak in the neutron spectrum. Altogether the neutron flux was conservatively assumed to be known with 10\% precision.

The MRPC detector to be tested (sec.\,\ref{sec:DetectorDescription}) was placed in the neutron beam at a distance of 11\,m from the lithium target. The electronics modules for the data acquisition were placed in the experimental cave at a distance of about 2\,m from the MRPC, but out of the neutron beam. 

\begin{figure}[]
\centering
\includegraphics[width=\columnwidth]{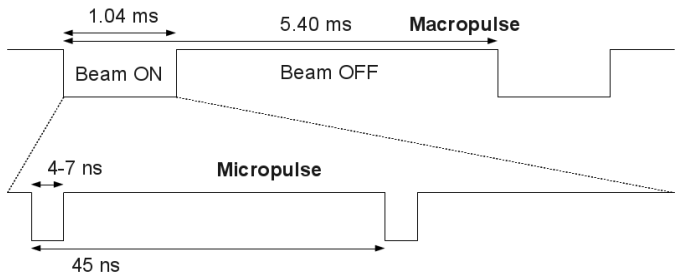}
\caption{Time structure of the beam-on and beam-off tags in the proton beam from the Gustav Werner cyclotron. The proton beam drives the neutron beam through the $^7$Li(p,n)$^7$Be reaction.}
\label{fig:Protonbeam}
\end{figure}

The proton beam from the Gustav Werner cyclotron had a time structure with a frequency of 185~Hz corresponding to a macropulse length of 5405\,$\mu$s (fig.\,\ref{fig:Protonbeam}). 

The beam was emitted only during the so-called beam-on time window, which was conservatively taken to be 1040\,$\mu$s long for the purposes of the data acquisition (DAQ). The offline analysis later showed that a somewhat shorter window of $\sim$850\,$\mu$s might have been chosen for the beam-on flag. As the flag affected the trigger condition, it was not possible to change the time interval flagged as beam-on in the offline analysis. During the beam-on time window, micropulses of 4-7 ns width and a repetition time of 45~ns were emitted. The number of protons emitted per micropulse varied strongly during the beam-on period. Depending on the precise run conditions, only for the first 810$\pm$40\,$\mu$s of the beam-on period protons were emitted by the cyclotron. 

The beginning of the acceleration was triggered by a pulse generator, and that signal was used as logical signal for the DAQ to trigger the beam-on range. During the acceleration, the cyclotron radiofrequency (RF) followed a pre-programmed time dependence. At the end of the acceleration cycle,
the RF system waited for the next trigger pulse from the generator.

%%%%%%%%%%%%%%%%%%%%%%%%%%%%%%%%%%%%%%%%%%%%%%%%%%%%%%%%%%%%%%%%%%%%%%%%%%%%%%%%%%%%%%%%
\section{Data acquisition}
\label{sec:DAQ}
%%%%%%%%%%%%%%%%%%%%%%%%%%%%%%%%%%%%%%%%%%%%%%%%%%%%%%%%%%%%%%%%%%%%%%%%%%%%%%%%%%%%%%%%

The electron avalanches in the MRPC gas gaps induced signals on the central anodes. These current signals were transported to
either FOPI~\cite{Ciobanu07-IEEE} or PADI~\cite{Ciobanu08-IEEE} front end electronics (FEE).
The FEEs had adjustable thresholds. They were located directly at each end of the strips and gave both a timing and a charge output. 

The timing output was connected to a 25\,ps multihit time-to-digital-converter (TDC, model CAEN V1290). The multihit TDC recorded all hits in a 200\,ns long search window, i.e. with a certain probability it recorded hits stemming from several cyclotron pulses that were just 45\,ns apart (sec.\,\ref{sec:TSL}). The charge output was amplified and, after amplification, fed into a charge-to-digital converter (QDC, model CAEN V965).

The logic for the trigger was implemented and controlled by a field programmable gate array (FPGA, model CAEN V1495) unit.
The FPGA also included programmable scalers that were driven by the TDC single trigger rates and the neutron flux monitors that were provided by the TSL facility (sec.\,\ref{sec:TSL}). The VME bus data acquisition was administered by a GSI multi-branch system (MBS)~\cite{Essel00-IEEE} device. 

Two different trigger conditions were used, depending on the macropulse signal provided by the accelerator: 
\begin{itemize}
\item[T1] When the beam was on (fig.\,\ref{fig:Protonbeam}), a trigger was issued when the RF signal from the accelerator coincided with valid time signals on both sides of any of the eight anode readout strips from the MRPC under study. In the logic, the RF signal was delayed in such a way that its start always determined the starting flank of the trigger. The typical  rate for trigger condition T1 was 16\,kHz.
\item[T2] When the beam was off (fig.\,\ref{fig:Protonbeam}), there were no RF signals from the accelerator, so trigger condition T1 could not be used as it would have prevented any data from being written. Therefore, in the beam-off range, a trigger was issued any time there were valid time signals on both sides of any of the eight anode readout strips from the MRPC under study. The typical rate was 4\,kHz.
\end{itemize}
The data were exported to GSI list mode files and backed up to external servers. They were subsequently converted to Root files for the offline analysis.

%%%%%%%%%%%%%%%%%%%%%%%%%%%%%%%%%%%%%%%%%%%%%%%%%%%%%%%%%%%%%%%%%%%%%%%%%%%%%%%%%%%%%%%%
\section{Data analysis and efficiency determination}
%%%%%%%%%%%%%%%%%%%%%%%%%%%%%%%%%%%%%%%%%%%%%%%%%%%%%%%%%%%%%%%%%%%%%%%%%%%%%%%%%%%%%%%%

\begin{figure}[]
\centering
\includegraphics[width=\columnwidth]{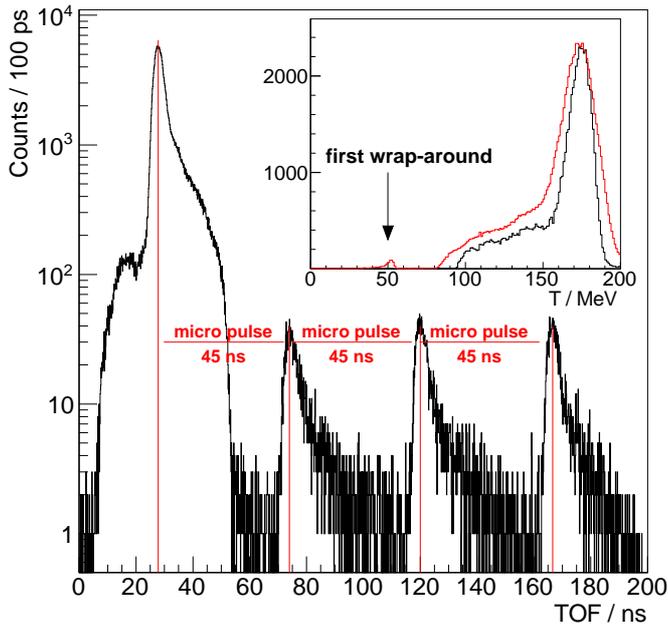}
\caption{Observed time of flight spectrum in the TDC window with beam-on. The structure between 0-60\,ns are the events that were included in the trigger. The structure then repeats itself every 45\,ns due to the repetition rate of the cyclotron, however with lower counting rate due to the efficiency of the detector. See text for a discussion of the inset. }
\label{fig:tof}
\end{figure}

When a valid trigger was issued, in a time window of 200\,ns all hits were accepted by the TDC without dead time (fig.\,\ref{fig:tof}). 

When the beam is on, the time-of-flight (TOF) spectrum starts with a plateau-like structure given by the neutron background that is not correlated to the beam. Subsequently, the typical structure of the TSL quasi-mono\-chro\-ma\-tic neutron beam becomes apparent, given by a strong peak near the nominal neutron energy followed by a shoulder at later times. 

The first structure in the time-of-flight spectrum has then been converted to kinetic energy (fig.\,\ref{fig:tof}, inset), assuming that the main peak is located at the nominal neutron energy as shown previously \cite{Bevilacqua11-NIMA}. The observed peak width is about 16\,MeV full width at half maximum at 175\,MeV, when the latter 200\,$\mu$s of the beam-on range are used for the conversion (fig.\,\ref{fig:tof}, thick black curve in the inset). Due to some jitter of the beam with respect to the RF signal issued to the DAQ, the peak is more smeared out, but still very apparent, when the full beam-on range is used (fig.\,\ref{fig:tof}, thin red curve in the inset). In the latter spectrum, also the peak due to the subsequent pulse, 45\,ns later, is drawn. This so-called first wrap-around is well separated and corresponds to 50\,MeV neutron kinetic energy for the present location, sufficiently low in energy that it does not disturb the analysis.

The low-energy shoulder to the left of the quasi-mo\-no\-chro\-ma\-tic peak stops at times equivalent to about 95\,MeV kinetic energy, after which only the non-correlated background remains. The lower edge of the spectrum was found to be slightly lower in the MEDLEY experiment, roughly 70-80\,MeV \cite{Bevilacqua11-NIMA}. This discrepancy may be explained by the fact that the efficiency for the MRPC-based detector is expected to rapidly drop at lower energies, due to the thick iron converters optimized for 200-1000\,MeV neutron kinetic energy.

% Reichweite von 100 MeV Protonen in Eisen: 11.3 g/cm2 [PSTAR], entspricht 2cm
% 40 MeV Protonen kommen nur 4mm weit in Eisen

The detection efficiency for neutrons was then determined using two different methods explained below. 

\subsection{Method I: Experimental dead time correction}

The first method, hereafter called method I, requires a determination of the effective time the DAQ is blocked after an event is recorded. Due to the large amount of data to be transferred (high-resolution TDC and QDC information for eight strips, usually two to three MRPCs that were placed one after the other in the neutron beam), it can be expected that the dead time is $>$90\%. 

\begin{figure}[]
\centering
\includegraphics[width=\columnwidth]{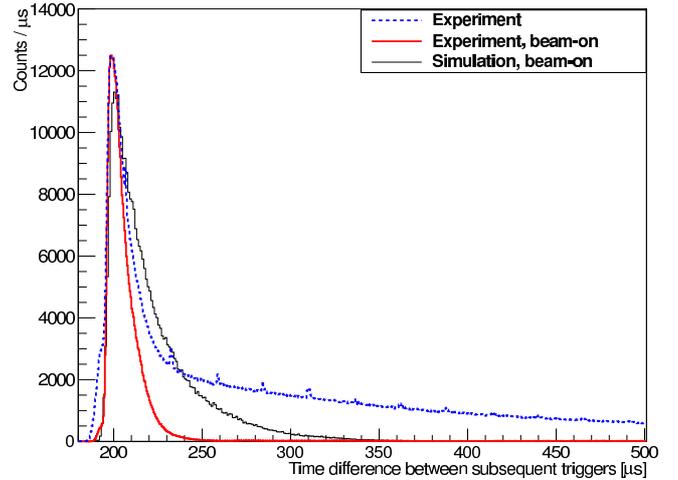}
\caption{Time difference between two consecutive triggers in a typical run with prototype HZDR-3c for the full data (dotted blue line) and restricted to the beam-on range (thick red line). A simulation assuming constant detection efficiency of 1.0\% and the observed time structure of the triggers (black curve in fig.\,\ref{fig:macro_meas_corr}), requiring both triggers to fall in the beam-on range is also shown (thin black line).}
\label{fig:TimeDifference}
\end{figure}

Indeed, a plot of the time difference between consecutive accepted triggers in the beam-on range shows a sharp peak at a time difference of 200\,$\mu$s, with almost no earlier events (fig.\,\ref{fig:TimeDifference}, blue dotted line). Using a Gaussian fit of this spectrum, a blocking time of 200$\pm$2\,$\mu$s is obtained and used henceforth. The shoulder to later times is due to background events which have somewhat lower efficiency and rate and therefore higher time difference between triggers. This shoulder is suppressed by requiring both triggers for the time difference to fall within the beam-on range (fig.\,\ref{fig:TimeDifference}, thick red line).

\begin{figure}[]
\centering
\includegraphics[width=\columnwidth]{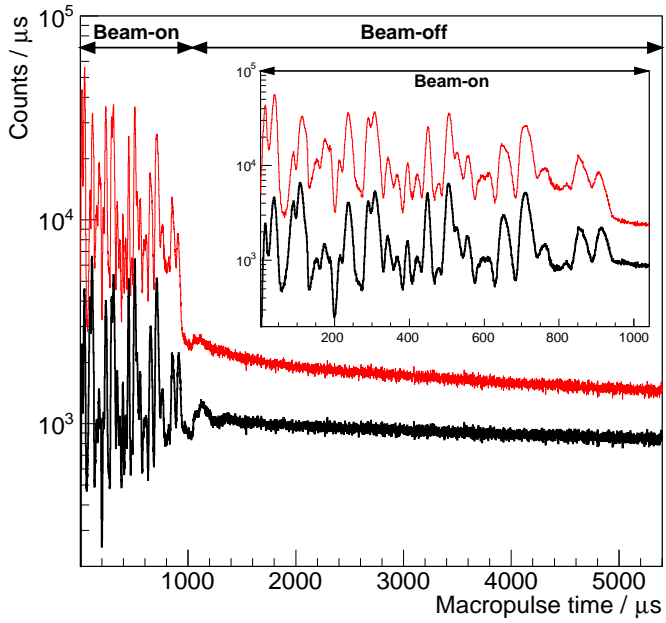}
\caption{Observed (thick black line) and dead-time corrected (thin red line) macropulse time spectra. See text for details.}
\label{fig:macro_meas_corr}
\end{figure}

Subsequently, the observed time spectrum given by the distribution of triggers during each macropulse (thick, black line in fig.~\ref{fig:macro_meas_corr}) is used in order to derive an experimental dead time correction. This is done in the manner described below.

When the beam is on (trigger condition T1), the spectrum shows no regular structure on the microsecond time scale (fig.~\ref{fig:macro_meas_corr}). This is due to the fact that the proton beam intensity of the cyclotron, and with it the neutron beam intensity, varies strongly and without regular structure during each macropulse. The regular 45\,ns micropulses are not resolved in this time scale. 

When the beam is off (trigger condition T2), a slowly decaying level of background is observed (fig.~\ref{fig:macro_meas_corr}). It has to be noted that the background level in the beam-off region is about 10\% higher than in the beam-on region, due to the more liberal trigger condition T2 applied here. Also, the long blocking time of 200\,$\mu$s (see above) leads to a high dead time of the system during the beam-on range, higher than in the beam-off range. These two effects raise the level of observed counts in the beam-off range with respect to the beam-on range.

\begin{figure}[]
\centering
\includegraphics[width=\columnwidth]{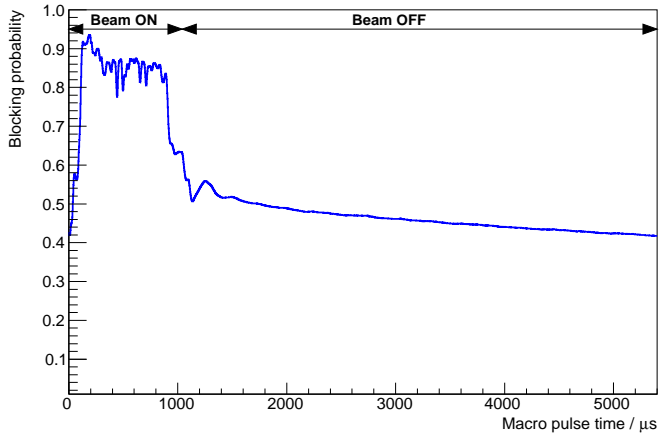}
\caption{Probability of blocking as a function of time in the macropulse. The curve is obtained based on the histogram of blocked bins, normalized to the total number of macropulses recorded for the run.}
\label{fig:Blocking}
\end{figure}

For the experimental dead-time correction, the above derived blocking time of 200$\pm$2\,$\mu$s is assigned to each observed trigger. In a new time histogram with 1~$\mu$s/channel dispersion, for each observed trigger the subsequent 200 channels are then incremented by one. In this way, a time spectrum of blocked events is created, reflecting the chan\-nel-dependent dead time. The more blocked events are found in one given channel, the higher the expected dead time for this channel. The time spectrum of blocked events is then divided by the total number of macropulses during the experimental run, giving the probability for an event to be blocked as a function of the time inside the macropulse (fig.\,\ref{fig:Blocking}).

This blocking probability, or channel-dependent dead time, is close to 90\% for most of the beam-on range. At the beginning of this range, the dead time is lower, because there the blocking is still mainly given by non-correlated events from the previous beam-off range. After the end of the beam-on range, the dead time remains high for some time due to the 200$\mu$s long blocking time and then drops slowly to about 50\%, the value given by the non-correlated events. 

%%%%%%%%%%%%%%%%%%%%%%%%%%%%%%%%%%%%%%%%%%%%%%%%%%%%%%%%%%%%%%%%%%%%%%%%%%%%%%%%%%%%%%%%

%%%%%%%%%%%%%%%%%%%%%%%%%%%%%%%%%%%%%%%%%%%%%%%%%%%%%%%%%%%%%%%%%%%%%%%%%%%%%%%%%%%%%%%%
%%%%%%%%%%%%%%%%%%%%%%%%%%%%%%%%%%%%%%%%%%%%%%%%%%%%%%%%%%%%%%%%%%%%%%%%%%%%%%%%%%%%%%%%

This channel-dependent dead time correction is then applied to the observed macropulse spectrum (fig.\,\ref{fig:macro_meas_corr}, thick black line) and results in a dead time-corrected macropulse spectrum (fig.\,\ref{fig:macro_meas_corr}, thin red line). From the beam-off range of this dead time-corrected spectrum, it is apparent that there is a background of about 2000 triggers/$\mu$s that is due to thermalized neutrons and, more importantly, $\gamma$ rays emitted during their capture. 

The inspection of the macropulse spectrum (fig.\,\ref{fig:macro_meas_corr}) between the effective end of neutron emission at 810$\pm$40\,$\mu$s and the end of the beam-on range at 1040\,$\mu$s shows that this background is also present during the beam-on range, with a level that is only about 10\% lower than when the beam is off despite the different trigger condition applied. Therefore, this background is subtracted here using the average trigger rate in the last 40\,$\mu$s of the beam-on range. This background amounts to roughly 20\% of the detected counts during the beam-on range. The efficiency $\eta({\rm I})$ is then obtained as follows:
\begin{equation}
\eta({\rm I}) = \frac{N_{\rm beam\,ON}^{\rm corr}-B_{\rm beam\,ON}^{\rm corr}}{\Phi_{\rm n} \, t \, A}
\end{equation}
with $N_{\rm beam\,ON}^{\rm corr}$ and $B_{\rm beam\,ON}^{\rm corr}$ the counts and the background in the corrected macropulse spectrum (fig.\,\ref{fig:macro_meas_corr}, thin red curve) in the beam-on range, $\Phi_{\rm n}$ the neutron flux as determined by the neutron monitors (sec.\,\ref{sec:TSL}), $t$ the running time and $A$ the area of the neutron beam spot at the position of the MRPC under study. 

The efficiency thus determined lies in the 1\% range (Table\,\ref{Table:Efficiency}). The systematic uncertainty is given by the uncertainty on $\Phi_{\rm n}$ (10\%) and on the dead time determination (10\%), altogether 14\%.

%%%%%%%%%%%%%%%%%%%%%%%%%%%%%%%%%%%%%%%%%%%%%%%%%%%%%%%%%%%%%%%%%%%%%%%%%%%%%%%%%%%%%%%%
\begin{table}
\centering
\begin{tabular}{llr@{$\pm$}l}
Prototype & Method & \multicolumn{2}{c}{Efficiency [\%]} \\ \hline
HZDR-1b & Experimental, I & 0.99 & 0.10$\pm$0.14 \\
 & Experimental, II & 1.00 & 0.15$\pm$0.11 \\ % Run 1360, Strip 4, Workbook p. 106 Analysis_Run1360_FZD1_strip4.pdf
 & Simulated & 0.90 & 0.01 \\
HZDR-3c & Experimental, I & 1.00 & 0.10$\pm$0.14 \\
 & Experimental, II & 1.00 & 0.15$\pm$0.11 \\
 & Simulated & 0.90 & 0.01 \\
\hline
\end{tabular}
\caption{\label{Table:Efficiency} Measured and simulated efficiency for the two MRPC prototypes under study here. The first error bar is statistical (dominated by background estimation and run-to-run reproducibility), the second error bar is the systematic uncertainty. The efficiency determination methods I and II are explained in the text.}
\end{table}

\subsection{Method II: Use of the second hit in the TDC search window}

An alternative method that does not depend on the dead time of the system takes advantage of the fact that the TDC accepts several hits within its search window. Therefore, once the DAQ is triggered by one hit in a TDC, all subsequent hits in this and all other channels are recorded without any dead time. So there may be altogether three or four hits in the TDC range (fig.\,\ref{fig:tof}). 

For the efficiency analysis, one requires a valid hit on both ends of one of the two central strips of the eight-strip MRPC structure and requires in addition that the weighted charge on that strip is higher than on any other strip, in order to exclude cross-talk events. Then, the peak near 75\,ns in fig.\,\ref{fig:tof} is integrated, both for the second hit histogram in the strip under study and in the full time spectrum of the other strips, giving $N_2$. This 75\,ns peak is recorded free of dead time, as discussed above. Then, $N_1$, the integral of the structure  from 0-60\,ns in fig.\,\ref{fig:tof}, is determined, quantifying the number of valid events used in the analysis. The efficiency $\eta$(II) is then obtained as follows:
\begin{equation}
\eta({\rm II}) = \frac{N_2 \,}{N_1 } \frac{185 \, \cdot \, 18000}{\Phi_{\rm n} \, A}
\end{equation}
with $\Phi_{\rm n}A/(185 \cdot 18000)$ the number of neutrons per micropulse for 185 macropulses per second and 18000 micropulses per macropulse. The results compare well to the data from method I (Table\,\ref{Table:Efficiency}). The systematic uncertainty for the efficiency determined by this method is given by the uncertainty on $\Phi_{\rm n}$ (10\%) and the uncertainty in the effective macropulse length (5\%), altogether 11\%.

%%%%%%%%%%%%%%%%%%%%%%%%%%%%%%%%%%%%%%%%%%%%%%%%%%%%%%%%%%%%%%%%%%%%%%%%%%%%%%%%%%%%%%%%
\section{Monte Carlo simulation}
%%%%%%%%%%%%%%%%%%%%%%%%%%%%%%%%%%%%%%%%%%%%%%%%%%%%%%%%%%%%%%%%%%%%%%%%%%%%%%%%%%%%%%%%

In order to perform a comparison to simulation,
the energy spectrum and the areal size of the neutron beam were coded in Geant4 version 9.4.p01.
The low energy neutron datafile \verb!G4NDL3.14! and the \verb!HadronPhysicsQGSP_BIC_HP! physics list
were used. The neutrons generated secondary particles, which were tracked by the Monte Carlo
engine. 

Three
parameters were applied: The threshold of the front-end electronics, a parameter characterizing the interplay between the avalanches created by the many
primary electrons in the gas, and a third one to handle the
space--charge effect. These parameters have been previously fixed by test experiments using an electron beam \cite{Elekes13-NIMA}
and were left untouched when simulating the present neutron beam data.

The simulation gave efficiency values of 0.90\% for both HZDR-1b and HZDR-3c prototypes. These numbers are in good agreement with the experimental data,
taking into account the systematic uncertainties (Table\,\ref{Table:Efficiency}), further corroborating the results of the simulation.

As already reported recently \cite{Elekes13-NIMA}, the small prototypes tested in the TSL experiment described here were scaled up to detectors of final size 200$\times$50~cm$^2$.
These very large MRPC structures were the building blocks of a possible high--efficiency array for fast neutron detection. This possible array has a layered structure, each with four MRPC elements \cite{Elekes13-NIMA}. In a Geant4 simulation, it was shown that
the requested efficiency for 200-1000\,MeV neutrons can be reached using 50 layers, and that a very good momentum resolution is achievable \cite{Elekes13-NIMA}. However, due to concerns about the multi-neutron capability, this design was not adopted for the NeuLAND detector, as discussed above in the introduction.  

Even though the present detector concept will not be applied at NeuLAND and FAIR, it is of relevance to understand which properties an MRPC-based neutron time of flight detector may have at lower energies, e.g.\ the radioactive ion beam facilities in operation or under construction at RIKEN (Wako, Japan) and at Michigan State University (MSU, East Lansing, USA). 

In order to address this question, the lower energy response (below 200~MeV) of the hypothetical array is investigated here, using the Monte Carlo code and modified ideal input described previously \cite{Elekes13-NIMA}. This ideal input assumes a breakup reaction $^{132}$Sn$\rightarrow$$^{132-x}$Sn+xn (x=1,2,3,4) at 50, 100, 150, and 200 MeV/A beam energies, respectively. The relative energies between the fragments were set to 100\,keV and 1000\,keV with a resolution of $\sigma$=0.
Eventfiles with the momenta of the neutrons were produced by the \verb!TGenPhaseSpace! class of ROOT~\cite{Root-Webpage}.
The neutrons were shot at the array from a distance of 12.5~m,
and 10000 events were processed for each scenario.
For the reactions when the number of emitted neutrons were more than one, the hit identification
algorithm detailed in Ref.~\cite{Elekes13-NIMA} was used.

For the 1n-scenario, an efficiency of 39\% was found at 50\,MeV/A, and higher values at higher beam energies (Table\,\ref{Table:Resolution}). The simulated resolution in the reconstructed relative energy spectra allows to study the astrophysically relevant range at 100\,keV relative energy with satisfactory efficiency and resolution (Table\,\ref{Table:Resolution}). 

However, as a consequence of the fact that the efficiency values for one neutron are well below 100\%, there is a low detection efficiency for multineutron scenarios. It should be noted here that the efficiency does not significantly improve when adding more layers to the 50-layer setup envisaged here. Rather, the 1n-efficiency is limited by the fact that at the present intermediate energies (50-200\,MeV/A) some of the charged particles produced by the neutrons might not reach the active gas layer and stop in the inactive steel and glass volumes of the MRPC structure. 

Therefore, an attempt was made to try to compensate the efficiency decrease by removing the steel converters from the array, as has been done previously for the MONA array at MSU \cite{Baumann05-NIMA}. This change in the setup resulted in an increase of the 1n detection efficiency at 50\,MeV/A from 39\% to 52\%, indicating that already the unavoidable glass volumes absorb many secondary charged particles. At higher beam energies, there was even no significant improvement in detection efficiency at all. This means that for multineutron scenarios, the setup cannot be efficiently used at low energies.
 
\begin{table}
\centering
\caption{Performance of a hypothetical detector array \cite{Elekes13-NIMA} consisting of 50 layers of the present MRPC structure: Monte Carlo simulated detection efficiency $\eta$ and resolution $\Delta E_{\rm rel}$ of the peak in the relative energy ($E_{\rm rel}$) spectra for the 1--neutron scenario at different beam energies.}
\begin{tabular}{r c c c c}
\hline\hline
\multicolumn{1}{c}{$E_{\rm beam}$} & Efficiency & \multicolumn{2}{c}{Resolution $\Delta E_{\rm rel}$} \\
(MeV/A) & $\eta$ & $E_{\rm rel}$=100\,keV & $E_{\rm rel}$=1000 keV \\
\hline
50 & 39\% & 11 & 43 \\
100 & 60\% & 16 & 48 \\
150 & 72\% & 16 & 48 \\
200 & 84\% & 23 & 89 \\
\hline
\end{tabular}
\label{Table:Resolution}
\end{table}

\section{Summary and conclusions}

A neutron time of flight detector consisting of passive steel converter and a multigap resistive plate chamber \cite{Yakorev11-NIMA} was tested using the 175\,MeV quasi-monochromatic neutron beam at TSL Uppsala. The detection efficiency was determined to be close to 1\% for one detector layer with two independent methods: The first one based on an experimental dead time correction, and the second one based on the second hit in multihit time-to-digital converters. 

The data are found to be in excellent agreement with a Monte Carlo simulation. This simulation had previously been developed and validated using a 31\,MeV electron beam \cite{Elekes13-NIMA}, and the present data offer important corroboration of the correctness of the simulation. 

The Monte Carlo simulation has then been used in order to derive the relevant characteristics of a hypothetical neutron-detector array made of 50\,layers of the present device. It was found that such a neutron detector array would work very well for scenarios where one neutron is to be detected, but only in a limited way for scenarios with several neutrons.

\section*{Acknowledgments}
This work was supported in
part by BMBF 06DR134I, BMBF 06KY7153, GSI F\&E (DR-ZUBE), NupNET NEDENSAA (BMBF 05 P 12 ODNUG), the European Union \linebreak (FP6-EFNUDAT), the Helmholtz Association Detector \linebreak Technology and Systems Platform (DTSP), and TÁMOP 4.2.4.A/2-11-1-2012-0001.

\end{document}